\documentclass[prl,twocolumn,superscriptaddress,showpacs]{revtex4}
\usepackage[english]{babel}
\usepackage[dvips]{graphicx}
\usepackage{psfrag}
\begin{document}
\title{Self-organization of irregular NEM vibrations in multi-mode
shuttle structures}
\author{L. M. Jonsson} \affiliation{Department of
Applied Physics, Chalmers University of Technology, SE - 412 96
G{\"o}teborg, Sweden} \author{F. Santandrea} \affiliation{Department
of Physics, University of Gothenburg, SE - 412 96 G{\"o}teborg, Sweden}
\author{L. Y. Gorelik} \email{gorelik@fy.chalmers.se}
\affiliation{Department of Applied Physics, Chalmers University of
Technology, SE - 412 96 G{\"o}teborg, Sweden} \author{R. I. Shekhter}
\affiliation{Department of Physics, University of Gothenburg, SE - 412
96 G{\"o}teborg, Sweden} \author{M. Jonson} \affiliation{Department of
Physics, University of Gothenburg, SE - 412 96 G{\"o}teborg, Sweden}
 \affiliation{School of Engineering and
Physical Sciences, Heriot-Watt University, Edinburgh EH14 4AS,
Scotland, UK}

\begin{abstract}

We investigate theoretically multi-mode electromechanical ``shuttle"
instabilities in DC voltage-biased nanoelectromechanical
single-electron tunneling (NEM-SET) devices. We show that initially
irregular (quasi-periodic) oscillations, that occur as a result of the 
simultaneous self-excitation of several mechanical modes with incommensurable
frequencies, self-organize into periodic oscillations with a frequency
corresponding to the eigenfrequency of one of the unstable modes. This 
effect demonstrates that a local probe can selectively excite global
vibrations of extended objects.
\end{abstract}


\maketitle

In a nanoelectromechanical single-electron tunneling (NEM-SET)
device mechanical vibrations and single-electron tunneling events
are coupled on the nanometer length scale. Ten years ago Gorelik {\em et al.} 
\cite{GorelikIsacsson98} suggested that such a coupling of mechanical 
and electrical degrees of freedom could lead to a ``shuttle instability'' 
and to a novel shuttle mechanism for charge transport through a DC 
voltage-biased NEM-SET. The theory \cite{GorelikIsacsson98} was developed 
for a rigid nanometer-size metal cluster suspended by elastic links between 
a source and a drain electrode so that only one mechanical degree of freedom 
was involved. Internal cluster vibrations were assumed to have much higher 
frequencies than the relatively low-frequency center-of-mass vibration 
and were therefore ignored. In the proposed shuttle regime charge 
transport between source and drain is mechanically assisted by pronounced
center-of-mass vibrations of the cluster (for reviews see 
\cite{ShekhterGalperin03,ShekhterGorelik06}).

\indent In this Letter --- motivated by the recent experimental discovery of
the shuttle instability predicted in Ref.~\cite{GorelikIsacsson98}
and of the related self excitation of radio-frequency mechanical
vibrations of gold-capped silicon nano-pillars \cite{KimQin07} ---
we generalize the theory of the shuttle phenomenon to include
extended structures with (many) internal mechanical degrees of
freedom. Our objective is to stimulate further experimental work on
NEM-SET systems involving, {\em e.g.}, suspended carbon nanotubes
and other extended molecules. In many such systems internal
mechanical degrees of freedom cannot be ignored and therefore a
theory of shuttling in the multi-mode regime needs to be developed.
This will be done in what follows.

Our main result is that when self excitation involves more than one
mode, strong mode-mode interactions caused by the non-linear coupling
of the mechanical vibrations to the electronic subsystem results in
self-organization of the vibrations. This implies that initially
quasi-periodic vibrations are transformed into pronounced periodic
vibrations as the shuttle instability develops towards a steady
state. The frequency of the steady-state vibrations depends crucially
on initial conditions and corresponds to the eigenfrequency of one of
the unstable modes.
\begin{figure}
\centering 
\includegraphics[width=0.8\linewidth]{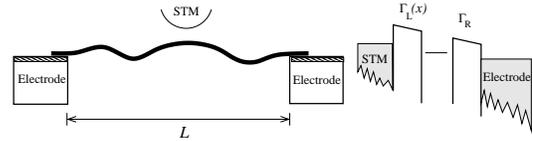}
\caption{Sketch of the model system considered. The electron tunneling
rate $\Gamma_{\rm L}(x)$ between the STM-tip and the center of a suspended 
carbon nanotube depends on the tube position, while the tunneling rate
$\Gamma_{\rm R}$ between tube and electrodes is constant. 
Tunneling may lead to self-excitation of multiple nanotube 
bending modes, which self-organize into pe\-rio\-dic, single-mode 
oscillations (the vibration amplitude shown is exaggerated).}
\label{fig:modelsystem}
\end{figure}

We will consider the particular realization of a multi-mode shuttle
structure sketched in Fig.~\ref{fig:modelsystem}. The sketch shows a
NEM-SET device based on a carbon nanotube (CNT) suspended over a
trench so that a segment of the tube is free to move in response to
external forces. Charge is injected into the suspended nanotube
through the tip of a scanning tunneling microscope (STM) positioned
above the center of the tube \cite{comment} as in the experiment of
LeRoy {\em et al.}  \cite{LeRoyLemay04}. For this very system a
current driven nanoelectromechanical instability has been predicted
and theoretically shown to result in the onset of pronounced CNT
bending mode vibrations involving one or more modes
\cite{JonssonGorelik05,JonssonGorelik07}. Below we will consider the
detailed development of the predicted instability when several bending
modes are unstable in order to find the resulting steady state. 
While a nanoelectromechanical instability involving these modes has not yet
been observed experimentally, bending mode vibrations of suspended CNTs 
have been detected in devices of the type studied here 
\cite{SazonovaYaish04,WitkampPoot06}. We note in passing that the 
electromechanical coupling to other types of CNT vibration modes have 
also been studied\cite{LeRoyLemay04,ZazunovFeinberg06,IzumidaGrifoni05,Flensberg06,SapmazJarilloHerrero06}.

We use continuum elasticity theory to describe the nanotube
mechanics \cite{WernerZwerger04,SapmazBlanter03}. The set of normal
modes for the bending mode vibrations is characterized by the
frequencies $\omega_n = \frac{c_n}{L^2}\sqrt{\frac{EI_z}{\rho S}}$,
where $E$ is Young's modulus of the CNT, $I_z$ is the area moment of inertia 
of the cross section, $\rho$ is the mass density, $S$ is the cross 
section area and the coefficients $c_n = 22.4, 61.7, 120.9,
199.9, 298.6, \dots$ are obtained by solving the equation
$\cos \sqrt{c_n} \cosh \sqrt{c_n} = 1$ \cite{LandauLifshitzElasticity}. 
The displacement $x(t)$ of the center of the CNT (see Fig.~\ref{fig:modelsystem}) 
can be expressed as a sum of normal-mode amplitudes $x_{n}(t)$, where only modes 
with $n=1,$ 3, 5, $\dots$ contribute due to symmetry. Other modes are 
inert and can be neglected. Each normal mode coordinate $x_{n}$ obeys Newton's 
equation in the form
\begin{equation}
\ddot x_n + \gamma \dot x_n + \omega_n^2 x_n = q\mathcal{E} / {M},
\label{eq:oscillators}
\end{equation}
where the force term depends on the average net charge $q$ on the CNT
and the effective electric field $\mathcal{E}$ that acts on the tube 
as a result of the applied bias voltage \cite{JonssonGorelik07}. 
The damping coefficient $\gamma$ in Eq. (\ref{eq:oscillators}) is a constant
in our model and therefore affects all modes in the same way. It is 
related to the quality factor of the vibration modes through the 
relation $Q_n= \omega_n / \gamma$. Experimentally $Q$ factors of order 1000
have been reported for nanoscale CNT based resonators in vacuum, 
\cite{HighQCNT}, while $Q$ factors as high as $10^4 - 10^5$ have been found 
for somewhat larger Si \cite{SiResonators06} and SiC resonators 
\cite{HighQSiC07}. 

An important parameter for the system under consideration 
is the ratio $\omega/\Gamma$ between the typical CNT vibration frequency 
$\omega$ and the characteristic rate of electron tunneling $\Gamma$. Having in
mind that the mechanical vibration frequency is extremely low on an
electronic scale ($\omega \sim 10^{8} - 10^{9}$ s$^{-1}$ for $L \simeq
1$ $\mu {\rm m}$) we consider here the case $\omega \ll \Gamma$. 
In addition we focus on the single-electron shuttling regime,
where only one electron can be accumulated on the CNT due to Coulomb
blockade of tunneling. Introducing the probability $p$ to find an
extra electron on the nanotube the average excess charge is $q=ep$
while the kinetic equation for the time evolution of the probability
$p$ can be written as
\begin{equation}
\dot p = -\Gamma (x) p  + \Gamma_{\rm L} (x).
\label{eq:charge}
\end{equation}
Here $\Gamma (x) = \Gamma_{\rm L} (x) + \Gamma_{\rm R}$ with $x=\Sigma_{n}
x_{n}$ while $\Gamma_{\rm L} (x) = \Gamma_0 \exp (-x/\lambda)$ is the rate
of electron tunneling across the STM-CNT junction, so that the 
typical rate of electron tunneling is $\Gamma \equiv \Gamma_0
+ \Gamma_R$. The characteristic length $\lambda$ is known as the tunneling length.

Equations~(\ref{eq:oscillators}) and (\ref{eq:charge}) describe the
coupled nanoelectromechanical dynamics of the CNT-based NEM-SET
device. In order to stay in the single-electron shuttling regime one has to apply a 
small enough bias voltage: 0.1 V is a typical value for a CNT of length 
$L\simeq 1 \,\mu$m. Therefore, it is reasonable to consider the limit of weak 
nanoelectromechanical coupling where the tunneling rates are only weakly 
modified by the electrostatic force induced by an excess charge $q=e$. This
condition holds for a small enough voltage bias $V$ (and a correspondingly 
small effective electric field $\mathcal{E}$) when the shift of the nanotube 
equilibrium position due to a single excess charge, $d_{n}=e\mathcal{E} /
(M \omega_{n}^{2})$, 
is small on the scale of the tunneling length, \emph{i.e.} when
$d_{n}/\lambda \ll 1$.  In this limit the onset of a shuttle
instability occurs independently in the different vibration modes
\cite{JonssonGorelik07} and if $\omega \ll\Gamma$ the instabilities
are soft \cite{IsacssonGorelik98}, {\em i.e.} the amplitude $A_{n}$ of
stationary shuttle vibrations goes to zero as $\mathcal{E}$ approaches
a critical field $\mathcal{E}_{c} = 4\Gamma \lambda \gamma M /e$ from
above. For small enough positive values of $\mathcal{E}-\mathcal{E}_{c}$ 
the vibration amplitudes $A_{n}$ will therefore be small compared to the 
tunneling length $\lambda$ and one may expand in the small parameter 
$A_{n}/\lambda$. By keeping third order terms one captures the most 
important non-linear effects.

Now we analyze Eqs.~(\ref{eq:oscillators}) and (\ref{eq:charge})
in the limits discussed above. A formal solution of Eq.~(\ref{eq:charge}) is
\begin{equation}
p(t) = \sum_{m=0}^\infty \left(-\Gamma(x)^{-1}\partial_t\right)^m \Gamma_{\rm L}(x)/\Gamma(x)\,
\end{equation}
which is a series expansion in the small parameter
$\omega/\Gamma$. If $\omega_n \ll \Gamma$ for all modes $n$ it is
sufficient to retain only the first order term and
substitute the truncated solution into Eq.~(\ref{eq:oscillators}).
After expanding to third order in the displacements $x_{n}$, the 
resulting non-linear equations for $x_{n}$ are solved by choosing 
the Ansatz: $x_{n}(t)= \lambda A_{n}(t)\sin(\omega_{n}t+\chi_{n}(t))$ 
and by then averaging over the fast oscillations \cite{Nayfeh93}. The
remaining equations describe the slow time variation of the amplitudes
$A_n$ and phases $\chi_n$ of the vibrations ($\dot{A}_n$, $\dot{\chi}_n \ll \omega_n$). 
Since the oscillator frequencies $\omega_n$ are incommensurable (see above) 
the amplitude and phase equations are decoupled in the limit considered here. 
Then the amplitude equations are
\begin{equation}
\dot A_n = \alpha_n A_n (\delta_n - A_n^2 - 2 \sum_{m \ne n} A_m^2),
\label{eq:amplitudes}
\end{equation}
\vspace{-3 mm}
with
\[
\delta_n = 16 \bigg(1  - \frac{4 \gamma \Gamma \lambda}{\omega_n^2
d_n} \bigg) + \mathcal{O}\bigg(\frac{\omega_n^2}{\Gamma^2}\bigg), \, \alpha_n = 
\frac{d_n\omega_n^2}{128 \lambda \Gamma}\bigg[1+ \mathcal{O} \bigg( 
\frac{\omega_n^2}{\Gamma^2}\bigg)\bigg] \nonumber
\]
One notes from the expression of $d_n$ that the product
$\omega_{n}^2 d_{n}$ does not depend on $n$ and that therefore
$\delta_n$ and $\alpha_n$ in Eq.~(\ref{eq:amplitudes}) do not depend
on $n$ to leading order in $\omega_n/\Gamma$. This will be used in
the following analysis and allows us to write $\alpha_n = \alpha$
and $\delta_n= \delta$. The relevance of corrections to these values 
will be discussed later.

\begin{figure}
\includegraphics[width=0.75\linewidth]{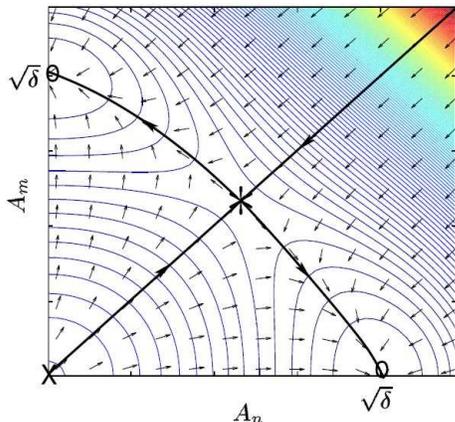}
\vspace{-5 mm}
\caption{(Color online) Stationary points when two modes ($n$ and
$m$) are unstable. Two attractors, indicated by ({\bf o}),
correspond to a finite amplitude of one mode while the other mode is
suppressed. The stationary point marked with ({\bf x}) is a repellor
and the point indicated by ({\bf *}) is a saddle point. The thick
lines are separatrices that trajectories cannot cross. The
separatrix $A_n = A_m$ ensures that if $A_n(0) > A_m(0)$, this
inequality hold for all times $t$.} \label{fig:trajectories}
\end{figure}

A complete analysis of how the solutions to Eq.~(\ref{eq:amplitudes}) 
evolve is possible when two modes $n,m$ are unstable. In this case 
one has two non-linear first order differential equations and the 
stationary points can be classified using standard techniques. The
stationary points found and their classification are listed in Table 
\ref{tab:classification} and the phase space is shown in Fig. 
\ref{fig:trajectories}. We find that the stationary solution corresponding 
to finite amplitude vibrations of both modes is a saddle point and the 
solutions describing finite amplitude vibrations of one mode and zero 
amplitude of the other are attractive fixed points. This means that 
depending on the initial conditions, one mode is selected to vibrate 
with finite amplitude while vibrations in the other mode are suppressed.
\begin{table}
\centering
\begin{tabular}{l|c}
 $\{A_1,A_3 \}$& Type \\ \hline $\{0 ,0 \}$ & Repulsive fixed point \\
 $\{0 ,\sqrt{\delta} \}$ & Attractive fixed point \\ $\{\sqrt{
 \delta} , 0 \}$ & Attractive fixed point \\ $\{ \sqrt{\delta/3} ,
 \sqrt{\delta/3} \}$ & Saddle point\\
\end{tabular}
\caption{Classification of stationary points when two modes ($n$=1
and 3) are included and both are unstable ($\delta >0$). Four
stationary points where $A_1,A_3 \ge 0$ are found. Two of them are
attractors and correspond to only one unstable mode, the other being
stable. Which mode is unstable is determined by the initial
conditions.} \label{tab:classification}
\end{table}
Although a complete analysis of the general non-linear problem can not
be done, a general statement for a system with $N>2$ modes can be
formulated for small $\delta$. In this limit one finds that the $N$
coupled non-linear equations (\ref{eq:amplitudes}) are characterized
by $N$ stable fixed points in the vector space
$\vec{A}=\{A_{n}\}$. The evolution of the system in this space is
represented by the motion of the point $\vec{A}(t)$ along a certain
trajectory. The stable points are given by $\vec{A}_{l}(\infty)=
(A_{l}(\infty)=\sqrt{\delta},\quad A_{n \neq l}(\infty)=0)$,
corresponding to one mode vibrating with amplitude $\sqrt{\delta}$ and
all other modes having zero amplitude. More precisely we have proven
the following theorem: if the initial conditions are such that
$A_{l}(0)>A_{n\neq l}(0)$, then the amplitude $A_{l}(t)$ increases
monotonically towards the final value $A_{l}(\infty)=\sqrt{\delta}$
while the other amplitudes $A_{n\neq l}(t)$ decay exponentially
towards zero as $t \rightarrow \infty$. The asymptotic estimate 
$A_{n\neq l}(t)\leq \sqrt{\delta}\exp\{-\alpha t(A_{l}^2(0)-A_{n}^2(0))\}$ 
gives an upper limit to the decaying amplitudes.
The theorem guarantees a remarkable feature of the shuttle
instability in that it allows a selective amplification of one of
the normal modes of the shuttle vibrations. Initially quasi-periodic
vibrations (for incommensurable frequencies $\omega_{n}$) are forced
to self-organize into regular large-amplitude vibrations with a
frequency corresponding to that of a single normal mode.
\begin{figure}[!t]
\includegraphics[width=0.95\linewidth,clip]{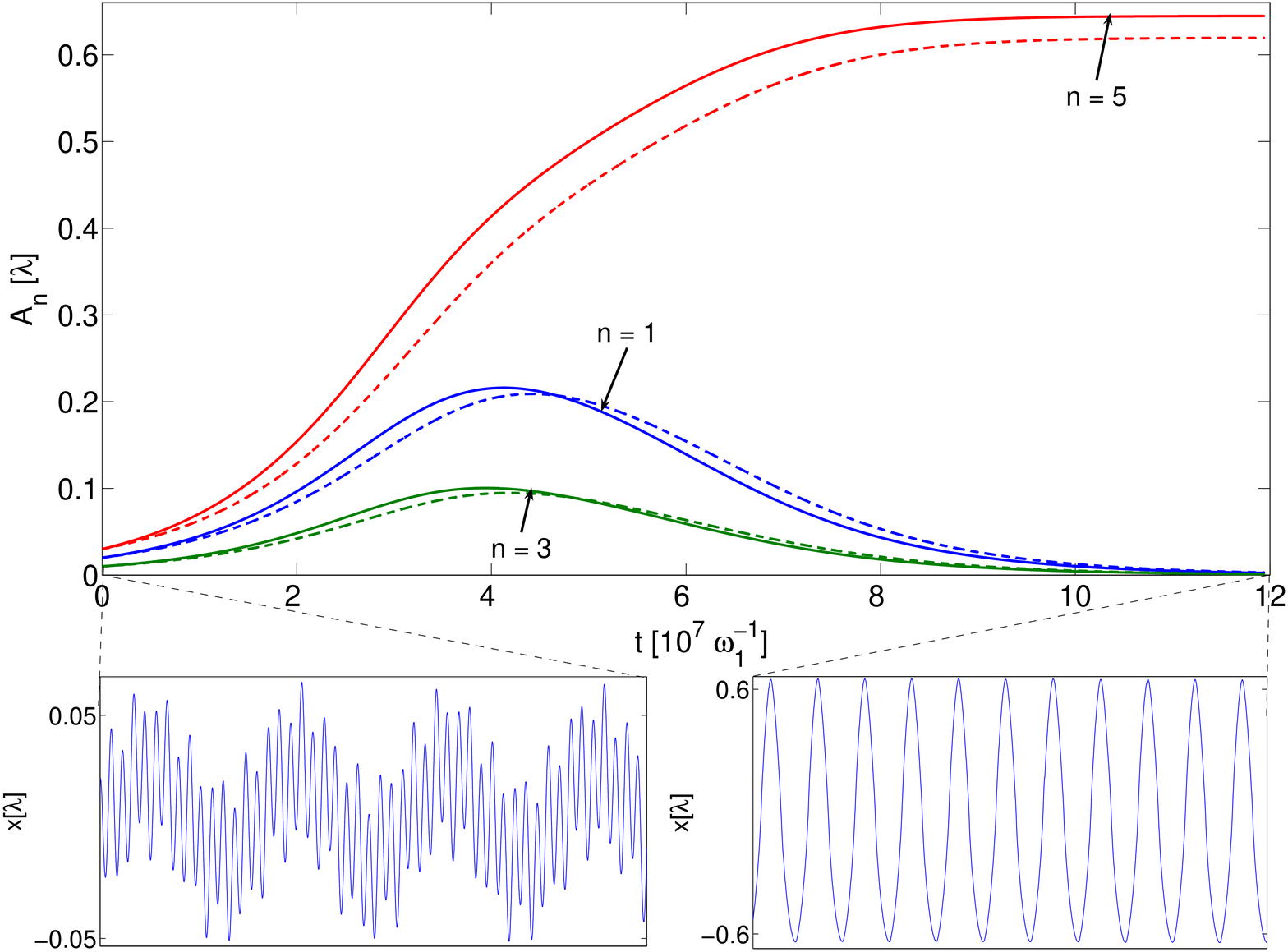}
\vspace{-5 mm}
\caption{(Color online) Nanotube vibration amplitude as a function of
time when three modes ($n=1,3,5$) are unstable ($\delta >
0$) and included in a full numerical solution of
Eqs.~(\ref{eq:oscillators})-(\ref{eq:charge}) with $ \gamma/\omega_1 = Q_1^{-1}=6.1 
\times 10^{-6},\, d_1/\lambda = 0.01$, $\omega_1/\Gamma = 0.0025$ and initial 
conditions $A_1(0) = 0.02$, $A_3(0) = 0.01$, $A_5(0) = 0.03$. These parameters were 
chosen to allow a comparison with the approximate result of Eq.~(\ref{eq:amplitudes}) 
which is plotted as dashed curves. The lower left panel shows the 
quasi-periodic oscillation of the nanotube center position just after the 
onset of the instability, while the lower right panel shows the regular
vibrations that appear after all but the n=5 mode amplitudes have been 
suppressed (see text).} \label{fig:amplitudes}
\end{figure}

We now resort to a numerical analysis in order to verify and
generalize the analytical results obtained above. We begin by using 
the same parameter range as before, \emph{i.e.} we choose parameters
in order to allow a comparison between the numerical and analytical 
results rather than to model realistic experiments. We will then 
show that the same type of behaviour obtains for more realistic sets
of parameters, for which no analytic solutions is available. 
Figure \ref{fig:amplitudes} shows numerical and analytical results for 
the time evolution of the vibration amplitude when there are three
different unstable mechanical modes ($n$=1, 3, 5). It is
clear that even though all three modes initially increase their amplitudes, 
a single mode is selected and the other modes are suppressed at large
times. Here the $n$=5 mode initially had the largest
deviation and hence the vibration frequency in the stationary state is
given by $\omega_5$. Similar figures when modes $n$=1 or 3 ends up
with a finite amplitude can be obtained by changing the initial
conditions. 

\begin{figure}
\includegraphics[width=0.85\linewidth,clip]{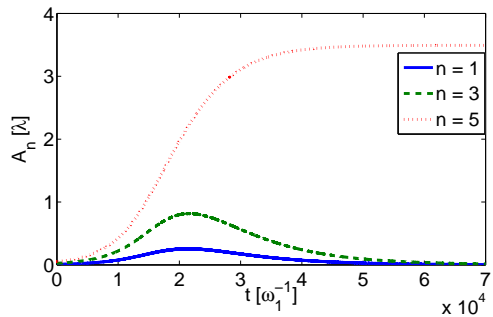}
\caption{(Color online) Time evolution of the vibration amplitude obtained from 
a numerical solution of Eqs.~(\ref{eq:oscillators})-(\ref{eq:charge}) when
three modes ($n=1,3,5$) are unstable and $\gamma / \omega_1=Q_1^{-1}= 5.5 \times
10^{-4}, d_1/\lambda = 0.7$ and $\omega_1/\Gamma= 0.005$. The initial conditions
were $A_1(0) = 0.005$, $A_3(0) =0.0025$ and $A_5(0) = 0.01$. The large amplitudes 
make an approximate analysis based on Eq.~(\ref{eq:amplitudes}) invalid, but the
the phenomenon of a selective excitation persists even in this case --- the mode
with the largest initial deviation is selected to end up with a large vibration 
amplitude. }
\label{fig:amplitudeslarge}
\end{figure}

So far we have only discussed cases when the final vibration amplitude is 
smaller than the tunneling length $\lambda$. Numerically, we have also 
investigated the opposite situation and found results that are qualitatively 
similar. As an example, a full numerical solution of Eqs.~(\ref{eq:oscillators})
and (\ref{eq:charge}) for a case when the final amplitude exceeds $\lambda$ is 
shown in Fig.~\ref{fig:amplitudeslarge}.

Finally we discuss the coefficients $\alpha_n$ and $\delta_n$. In
the previous analysis we used the approximation that these
coefficients are the same for all modes, which is only valid when
$\Gamma \gg \max \omega_n$ and the dissipation mechanism affects all the modes
in the same way. 
In general if one takes corrections of order $\omega_n^2/\Gamma_n^2$ 
into account or considers a mode-dependent dissipative mechanism, one has 
$\delta_n \ne \delta_m$. This breaks the symmetry between the modes and leads to
a situation where the final vibration state is not simply determined by the 
largest initial vibration mode amplitude. In this case the number of stable 
attractors can be smaller than the number of unstable modes. This implies that 
in some cases, even though a mode is initially unstable, it cannot reach finite 
amplitude vibrations since it is suppressed by another mode.

In conclusion we have investigated electromechanical instabilities
of different mechanical vibration modes of a suspended carbon
nanotube. We have shown that the excitation mechanism considered
leads to a selective excitation of a specific mode depending on the
initial conditions. This demonstrates a way of using local tunneling
injection of charge to probe the mechanics of extended nano-objects.
The analysis presented here can be generalized to apply to other
multi-mode shuttle structures. The phenomenon of a selective
excitation of a specific mode is general in the sense that even
though several modes are unstable, only a single mode reaches a
steady state with a finite amplitude.

Stimulating discussions with A. Isacsson and financial support from
the Swedish VR and SSF and from the EC (FP6-003673, CANEL) is gratefully
acknowledged.


\begin{thebibliography}{22}
\bibitem{comment}
 This configuration is chosen for simplicity. Our results are valid
 if the distance between the STM tip and the nanotube mid-point
 is small compared to the tube length.
\end{thebibliography}
\end{document}